\documentclass[twocolumn]{aastex62}
\usepackage{natbib}
\submitjournal{AJ (in press)}

\begin{document}

                                                                                                                                                               
\title{Chandra Resolves the Double FU Orionis System RNO 1B/1C in X-rays}

\correspondingauthor{Stephen L. Skinner}
\email{stephen.skinner@colorado.edu, manuel.guedel@univie.ac.at}

\author[0000-0002-3025-3055]{Stephen L. Skinner}
\affiliation{Center for Astrophysics and
Space Astronomy (CASA), Univ. of Colorado,
Boulder, CO, USA 80309-0389}

\author{Manuel  G\"{u}del}
\affiliation{Dept. of Astrophysics, Univ. of Vienna,
T\"{u}rkenschanzstr. 17,  A-1180 Vienna, Austria}

                                                                                                                                                               
\newcommand{\ltsimeq}{\raisebox{-0.6ex}{$\,\stackrel{\raisebox{-.2ex}
{$\textstyle<$}}{\sim}\,$}}
\newcommand{\gtsimeq}{\raisebox{-0.6ex}{$\,\stackrel{\raisebox{-.2ex}
{$\textstyle>$}}{\sim}\,$}}

\begin{abstract}
\small{We present new {\em Chandra} X-ray observations of the 
close pair of young stars RNO 1B and 1C (6$''$ separation)
located in the L1287 cloud. RNO 1B erupted in 1978 - 1990 and is 
classified as an FU Orionis star (FUor). RNO 1C also shows most
of the properties of an FUor but no eruption has yet been seen.
Only a few dozen FUors are known and the presence of two such 
objects with a small angular separation is rare, suggesting a 
common origin.
Both stars were faintly detected by {\em Chandra} and we
summarize their X-ray properties within the framework
of other previously detected FUors. We also report other
X-ray detections in L1287 including the deeply-embedded young star
RNO 1G, the jet-like radio source VLA 3, and an
enigmatic hard flaring source with no 2MASS counterpart
that was only detected in the second of two {\em Chandra} exposures.
}
\end{abstract}

\keywords{stars: individual (RNO 1B/1C) --- stars: pre-main-sequence --- X-rays: stars}

\section{Introduction}
FUors are low-mass pre-main sequence (PMS) stars which undergo large optical or
IR outbursts of several magnitudes followed by a slow decay on  timescales of
decades or longer. The prototype FU Ori erupted optically in 1936-37 and is still in
slow decline. The outbursts are thought to be due to a dramatic increase in the
accretion rate from the circumstellar disk onto the star. The enhanced accretion
is accompanied by development of a strong cool wind.
FUors also show  wavelength-dependent spectral
types mimicing low surface gravity F-G giants or supergiants in the optical
(attributed to a self-luminous accretion disk), infrared excesses, and 
2.3 $\mu$m CO absorption. 

Only about two dozen FUors are known (Audard et al. 2014; 
Connelley \& Reipurth 2018, hereafter CR18). It is thus
quite remarkable that a close pair of such objects,
RNO 1B and 1C (6$''$ separation) exists in the L1287 cloud.
At the L1287 distance of 929$\pm$34 pc (Reid et al. 2014),
their separation is $\approx$5574 AU.
The two stars probably  formed contemporaneously and it has been 
suggested that both 1B and 1C may be close binaries, comprising
a hierarchical quadruple system (Reipurth \& Aspin 2004).

The L1287 region contains RNO 1, which was listed in the catalog of
red and/or nebulous objects (RNO) compiled by Cohen (1980). 
Subsequently, Staude \& Neckel (1991, hereafter SN91) 
obtained $I$-band images which resolved the nebulous RNO 1 region 
into several compact knots including the close pair RNO 1B and 1C,
now known to be young stars (Kenyon et al. 1993).
They have similar 2MASS K$_{s}$ magnitudes of 
K$_{s}$ = 7.76 (1B) and K$_{s}$ = 7.54 (1C). Both are viewed
through high extinction. Estimates for RNO 1B are
A$_{\rm v}$ $\approx$ 9.2 mag (SN91) to 14.5$\pm$1 mag (CR18).
RNO 1C is redder with A$_{\rm v}$ $\approx$ 12 mag (SN91) to
19.5$\pm$4 mag (CR18). 
RNO 1B (= V710 Cas) brightened by at least 3 mag in $R$ band 
between 1978 - 1990 (SN91) and is a classical (outburst) FUor. 
No eruption has  yet been seen in RNO 1C but it does show most of the 
defining features of the class, so it is classified 
as FUor-like (CR18; Kenyon et al. 1993). 

We report the first X-ray detection of RNO 1B and 1C.
Our main objective was to use {\em Chandra}'s
excellent spatial resolution to resolve the pair and quantify their X-ray properties,
placing them into context with other X-ray detected FUors such as
FU Ori (Skinner et al. 2006; 2010) and the classical FUor
V1735 Cyg (Skinner et al. 2009).

\section{Observations}
We observed RNO 1B/1C with the {\em Chandra} Advanced CCD Imaging Spectrometer
(ACIS-I)  in two separate exposures
on 2018 Mar 20 (ObsId 20135; 34.507 ks livetime) and 
2018 May 29 (ObsId 21041; 24.636 ks) providing a total livetime of 59.143 ks.
The ACIS detector is sensitive in
the E  $\approx$0.4-10 keV energy range with a pixel size of 0.$''$492.
For on-axis sources, 90\% of the encircled energy fraction (EEF) lies 
within a radius of $R_{90}$ $\approx$0$''$.9 at E=1 keV.
The EEF is energy-dependent and $R_{90}$ increases toward higher 
energies\footnote{Details on {\em Chandra} instrumentation and performance can be
found in the {\em Proposer's Observatory Guide} at
cxc.harvard.edu/proposer/POG~.}.

Data were reduced using standard
threads in the {\em Chandra} Interactive Analysis of Observations (CIAO v4.11)
package. Events from the two exposures were reprojected onto the
same tangent point and merged using the CIAO  $merge\_obs$ script. 
The image created from the merged events is shown in Figure 1. 
For sources of interest, spectra and associated response  files 
were  extracted separately for each observation and then fitted simultaneously 
using {\em XSPEC} v. 12.10.1.   

\begin{figure}
\figurenum{1}
\includegraphics*[width=6.0cm,angle=0]{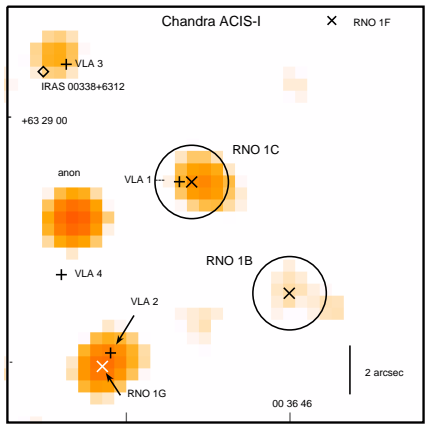} 
\caption{{\em Chandra} ACIS-I image (0.5-7 keV; log intensity scale) of the 
         region near RNO 1B/C based on merged event data from
         ObsIds 20135 and 21041 (59.143 ks).
         The image has been lightly Gaussian smoothed using a 3-pixel kernel
         to bring out faint emission from RNO 1B (5 counts) and 1C (8 counts).
         The 2MASS positions ($\times$) and $r$=1$''$.5 source extraction
         regions of RNO 1B and 1C are shown.
         The position of RNO 1G ($\times$) is from Quanz et al. (2007).
         VLA radio positions ($+$) are from Anglada et al. (1994).
         The source labeled ``anon'' is CXO J003647.3$+$632856.
         Sources RNO 1F and VLA 4 were not detected by {\em Chandra}.
}
\end{figure}

\begin{figure}
\figurenum{2}
\includegraphics*[height=6.0cm,angle=-90]{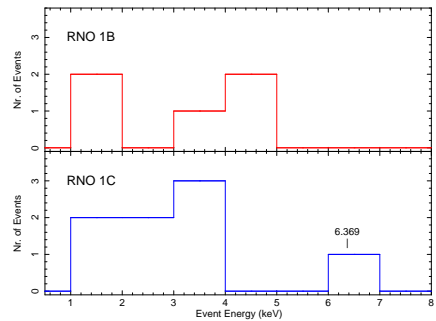} \\
\caption{X-ray event energy histograms for RNO 1B/1C using
energy bin widths of 1 keV.
}
\end{figure}

\section{Results}

\subsection{RNO 1B and 1C}
Table 1 summarizes the X-ray properties of RNO 1B and 1C and
other sources in their vicinity.
For RNO 1B, the summed exposures  yielded 5 events
inside a $r$=1$''$.5 extraction circle centered on its 2MASS position 
(2MASS J00364599$+$6328529). Three events were detected in the
first observation and two in the second with energies in the range
$E$ = 1.39 - 4.75 keV and mean energy $\overline{E}$ = 3.09$\pm$1.40 keV.
For RNO 1C (2MASS J00364659$+$6328574), a similar extraction yielded 8 events,
with four detected in each observation and a range $E$ = 1.45 - 6.37 keV
and $\overline{E}$ = 3.28$\pm$1.39 keV. Its X-ray centroid lies  0$''$.7
west of the radio source VLA 1 (J003646.67$+$632853.7; Anglada et al. 1994).
Background is negligible ($<$1 count within $r$=1$''$.5 extraction regions).

Event energy distributions are plotted as histograms in Figure 2.
The outlier event at 6.37 keV for RNO 1C is noteworthy 
since it may be due to fluorescent Fe produced when cold gas near 
the star is irradiated by hard X-rays, as has been detected in FU Ori 
itself (Skinner et al. 2006). But a spectrum of RNO 1C with more 
counts is needed to determine if fluorescent Fe emission near
6.4 keV is actually present.

We fitted the unbinned RNO 1C spectra
with a simple absorbed single temperature thermal plasma model (1T APEC) using
the C-statistic.  Since the two spectra together provide only 8 counts, they do not reliably
constrain the absorption column density N$_{\rm H}$. It was thus stepped through
values in the range N$_{\rm H}$ = (1.7 - 4.5)$\times$10$^{22}$ cm$^{-2}$, corresponding 
to A$_{\rm v}$ $\approx$ 9 - 24 mag in order to span previous extinction estimates. 
We adopt the conversion N$_{\rm H}$ = 1.9$\times$10$^{21}$A$_{\rm v}$ cm$^{-2}$ obtained
by averaging the slightly different conversions of Gorenstein (1975) and 
Vuong et al. (2003). The fits gave  observed (absorbed) fluxes
F$_{x,abs}$(0.3-8 keV) = 2.50($\pm$0.20)$\times$10$^{-15}$ ergs cm$^{-2}$ s$^{-1}$. 
The derived plasma temperature, unabsorbed flux, and intrinsic X-ray luminosity (L$_{x}$) are 
sensitive to the assumed value of N$_{\rm H}$. The C-statistic is minimized for
N$_{\rm H}$ = (2.4 - 2.6)$\times$10$^{22}$ cm$^{-2}$ (A$_{\rm v}$ $\approx$ 12.6 - 13.7 mag)
and kT $\approx$ 3.2 - 3.7 keV. The luminosity estimate is 
log L$_{x}$(0.3-8 keV) = 29.83($\pm$0.15) ergs s$^{-1}$.

Since only 5 counts were detected for RNO 1B we did not attempt spectral fits.
Instead, we used the Portable Interactive Multi-Mission Simulator (PIMMS)
to estimate the flux and L$_{x}$ based on the ACIS-I count rate of 
0.0845 c ks$^{-1}$. 
A 1T APEC absorbed thermal plasma model was used, as for RNO 1C. The
absorption was stepped through values in the range 
N$_{\rm H}$ = (1.7 - 3.0)$\times$10$^{22}$ cm$^{-2}$, corresponding
A$_{\rm v}$ $\approx$ 9 - 16 mag (Staude \& Neckel 1991; CR18).
We used two different plasma temperatures kT = 3 and 5.4 keV,
typical of FUors and classical (accreting) T Tauri stars.
For the above range of values, PIMMS predicts 
F$_{x,abs}$(0.3-8 keV) = 1.60($\pm$0.20)$\times$10$^{-15}$ ergs cm$^{-2}$ s$^{-1}$
and unabsorbed luminosity log L$_{x}$(0.3-8 keV) = 29.64($\pm$0.15) ergs s$^{-1}$.
We emphasize that this L$_{x}$ estimate is based on assumed values of
N$_{\rm H}$ and kT that lie within reasonable ranges, but are not 
actual measurements. If N$_{\rm H}$ is greater than assumed (as could occur
from cold gas along the line-of-sight that is not accounted for by A$_{\rm v}$) 
or if kT is less,  then higher L$_{x}$ values are possible.

\begin{widetext}
\begin{center}
\begin{deluxetable}{llcccclc}
\tabletypesize{\scriptsize}
\tablewidth{0pt}
\tablecaption{Chandra X-Ray Sources Near RNO 1B/C}
\tablehead{
           \colhead{Chandra position}               &
           \colhead{Name}               &
           \colhead{Net Counts}         &
           \colhead{$\overline{E}$ ($E_{50}$)}  &
           \colhead{Hardness}           &
           \colhead{F$_{x,abs}$}        &
           \colhead{log L$_{x}$}        &
           \colhead{Offset}     \\
           \colhead{R.A., decl. (J2000)}     &
           \colhead{}            &
           \colhead{(cts)}              &
           \colhead{(keV)}              &
           \colhead{}                   &
           \colhead{(ergs cm$^{-2}$ s$^{-1}$)}   &
           \colhead{(ergs s$^{-1}$)}           &
           \colhead{(arcsec)}
                                  }
\startdata
J00 36 46.04+63 28 52.8 & RNO 1B  & 5$\pm$2 & 3.08 (3.09) & 0.60  & 1.60($\pm$0.20)e-15\tablenotemark{b} & 29.64\tablenotemark{b}  &  0.34   \\
J00 36 46.55+63 28 57.3 & RNO 1C  & 8$\pm$3 & 3.28 (3.21) & 0.75  & 2.50($\pm$0.20)e-15\tablenotemark{c} & 29.83\tablenotemark{c}  &  0.31   \\
J00 36 47.12+63 28 50.1 & RNO 1G  & 9$\pm$3 & 4.71 (4.59) & 1.00  & 5.83($\pm$0.20)e-15                  & 30.41                   &  0.22   \\
J00 36 47.41+63 29 02.7 & VLA 3   & 4$\pm$2 & 5.42 (5.36) & 1.00  & ...\tablenotemark{d}     & ...\tablenotemark{d}                &  0.52   \\
J00 36 47.30+63 28 56.1 & anon    &10$\pm$3 & 4.58 (3.85) & 1.00  & 9.88($\pm$0.62)e-15\tablenotemark{e} & 31.04                   &  ...   \\
\enddata
\tablenotetext{a}{
Notes:~Data are based on merged events (0.3 - 7 keV)  from {\em Chandra} ObsIds 20135 and 21041,
with a total livetime of 59.143 ks. 
Tabulated quantities are: J2000.0 X-ray centroid position (R.A., decl.); object name; net counts and
net counts error; mean ($\overline{E}$) and median ($E_{50}$) event energies; 
Hardness =  counts(2-7 keV)/counts(0.3-7 keV);
absorbed X-ray flux (0.3-7 keV); unabsorbed X-ray luminosity at an assumed distance of 930 pc,
and offset between {\em Chandra} and IR or VLA radio positions.
}
\tablenotetext{b}{F$_{x}$ and L$_{x}$ are estimated from 1T APEC PIMMS simulations assuming absorption
                  N$_{\rm H}$ = (1.7 - 3.0) $\times$ 10$^{22}$ cm$^{-2}$ and kT = 3.0 - 5.4 keV.}
\tablenotetext{c}{F$_{x}$ and L$_{x}$ are estimated from 1T APEC fits of unbinned spectra assuming
                  N$_{\rm H}$ = (1.7 - 4.5) $\times$ 10$^{22}$ cm$^{-2}$.}
\tablenotetext{d}{Insufficient counts to determine F$_{x}$ and L$_{x}$.}
\tablenotetext{e}{F$_{x}$ is based on a 1T APEC fits of unbinned spectra with 
                 N$_{\rm H}$ = (1-2)  $\times$ 10$^{23}$ cm$^{-2}$ and kT = 1.6 - 4.1 keV.} 
\end{deluxetable}

\end{center}
\end{widetext}

\subsection{Other X-ray Sources Near RNO 1B/1C}
\noindent \underline{CXO J003647.1$+$632850.1}~is a 9-count X-ray source
whose position is nearly coincident with the infrared source RNO 1G 
(J003647.14$+$632849.95; Quanz et al. 2007). RNO 1G is
a deeply embedded young stellar object (YSO) originally identified in near-IR
polarization maps by Weintraub \& Kastner (1993). Both the X-ray and IR 
positions of RNO 1G  are offset by $\approx$0$''$.5  SE of radio continuum 
source VLA 2 (Anglada et al. 1994).  The X-ray source is hard
with event energies in the range $E$ = 3.46 - 6.19 keV.
Fits of the unbinned spectra with a 1T $APEC$ model give a minimum
C-statistic for N$_{\rm H}$ = (1.5 - 2.0)$\times$10$^{23}$ cm$^{-2}$, corresponding
to A$_{\rm v}$ $\gtsimeq$ 79 mag, and high plasma temperatures
kT $\approx$ 13 - 14 keV. 
\smallskip

\noindent \underline{CXO J003647.4$+$632902.7} is a faint source offset $\approx$0.$''$5
NE of the 3.6 cm radio continuum source VLA 3. The radio source shows
elongated structure and has a positive spectral index (Anglada et al. 1994),
as commonly observed for driving sources of bipolar outflows. It was thus
proposed by Anglada et al. that VLA 3 drives the bipolar molecular outflow
in L1287 that was studied in millimeter molecular lines by Yang et al. (1991).
There is no cataloged 2MASS source within 1$''$ of the X-ray position but
IRAS 00338$+$6312 lies $\approx$0$''$.9 SE of the X-ray centroid.
Four events were detected, of which 3 were
detected in the first observation. The source is very hard with event
energies in the range $E$ = 4.36 - 6.60 keV. The {\em Chandra}
X-ray position may point to a deeply embedded (proto)star associated
with the jet-like source VLA 3.

\smallskip
\noindent \underline{CXO J003647.3$+$632856.1} is a variable source which
remarkably was only detected in the second {\em Chandra} observation (10 counts). 
This source is labeled ``anon'' in Figure 1. There is no catalogued 2MASS counterpart. 
The radio source VLA 4 (J003647.39$+$632853.7) lies 2$''$.3 to the south.
The X-ray source is  hard with event energies $E$ = 2.93 - 6.82 keV.
Three events have energies E = 6.36 - 6.82 keV, suggestive of Fe emission.
This object is likely a heavily embedded magnetically active (flaring) young 
star or protostar that is only intermittently detected in X-rays.
We fitted the unbinned spectrum with a 1T APEC model. No {\em a priori}
information on A$_{\rm v}$ is available but we assumed high X-ray absorption
and stepped through values N$_{\rm H}$ = (1 - 40) $\times$ 10$^{22}$ cm$^{-2}$.
The C-statistic is minimized for N$_{\rm H}$ = (1-2)  $\times$ 10$^{23}$ cm$^{-2}$
(A$_{\rm v}$ $\gtsimeq$ 52 mag) with kT $\approx$ 2-4  keV.   

\smallskip
\noindent \underline{Non-Detections}~We find no significant X-ray emission at
the position of the source tentatively identified as RNO 1D by
Weintraub et al. (1996) or at the position of  RNO 1F (Quanz et al. 2007).

\begin{figure}
\figurenum{3}
\includegraphics*[height=7.0cm,angle=-90]{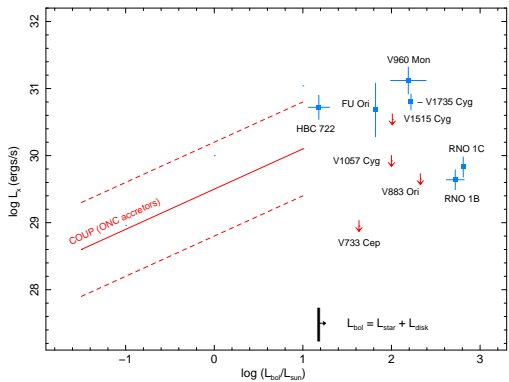} \\
\caption{X-ray luminosity (L$_{x}$) versus bolometric luminosity (L$_{bol}$/L$_{\odot}$)
for FUors  observed by {\em Chandra} and {\em XMM-Newton} based on published and
archival data. Downward arrows show upper limits for three FUors 
undetected by {\em XMM-Newton} (V1515 Cyg, V1057 Cyg, V883 Ori) and
for the {\em Chandra} non-detection of V733 Cep.
The regression fit for accreting sources in
the Orion Nebula Cluster (ONC) are shown (red solid line) along with 
the $\pm$0.7 dex (1$\sigma$) dispersion in log L$_{x}$ (red dashed line; Preibisch et al. 2005).
The high L$_{bol}$ values of FUors are dominated by the luminous accretion disk.
}
\end{figure}

\begin{figure}
\figurenum{4}
\includegraphics*[height=7.0cm,angle=-90]{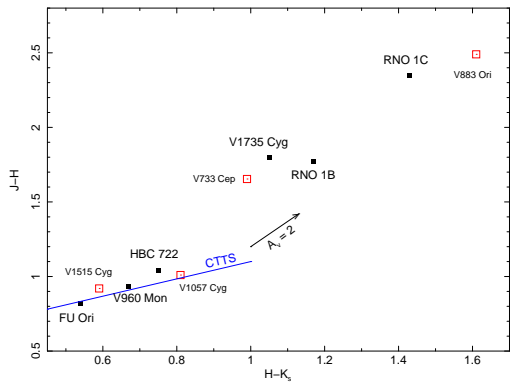} \\
\caption{2MASS near-IR colors for selected FUors  observed by {\em Chandra} and 
{\em XMM-Newton}. Solid squares are X-ray detections and open squares are
non-detections (see Fig. 3). The solid line showing {\em unreddened} colors
for classical T Tauri stars and the reddening vector for A${\rm v}$ = 2 mag
are based on Meyer et al. (1997).
}
\end{figure}

\section{Discussion}

\subsection{RNO 1B and 1C in Context}

The {\em Chandra} data show similar X-ray properties for RNO 1B and 1C.
There is no significant difference in the number of events detected for
the two stars or in their mean event energies to within the uncertainties.
Spectra with a larger number of counts will be needed to determine if 
real differences exist. Since no pre-outburst X-ray spectra are available
for RNO 1B (or for FUors in general) we do not know whether its X-ray 
properties were affected by the 1978 - 1990 eruption. 

Our L$_{x}$ estimates for RNO 1B and 1C make them the least
X-ray luminous FUors detected to date, about a factor of ten
less luminous than the X-ray bright sources 
FU Ori (log L$_{x}$ = 30.68 ergs s$^{-1}$), 
V1735 Cyg (log L$_{x}$ = 30.80), and V960 Mon (log L$_{x}$ = 31.12). 
The faint X-ray emitting FUor-like 
source L1551 IRS 5 has a lower L$_{x}$ than RNO 1B and 1C but its 
X-ray emission is  offset from the obscured central object and 
probably originates  in its jet 
(Favata et al. 2002; Bally et al. 2003; Schneider, G\"{u}nther, \& Schmitt 2011). 
Upper limits for a few undetected FUors
are comparable to or less than the L$_{x}$ values of RNO 1B and 1C. 
For example, V883 Ori was undetected by {\em XMM-Newton} at 
log L$_{x}$ $\leq$ 29.65 ergs s$^{-1}$ (ObsId 0205150501; PI: S. Skinner).
Also, three archived {\em Chandra} ACIS-I exposures with a combined livetime
of 71.675 ks (ObsIds 9919, 10811, 10812; PI: T. Allen) captured 
V733 Cep far off-axis but it was not detected. This star erupted
in 1971 or earlier and has optical and near-IR spectra  very
similar to FU Ori (CR18). Based on all three {\em Chandra} exposures,
we obtain a conservative upper limit log L$_{x}$(0.3-8 keV) $\leq$ 28.95 ergs s$^{-1}$
at d = 800 pc (CR18) assuming a thermal plasma spectrum with kT = 3 keV
and extinction as large as A$_{\rm v}$ = 11.5$\pm$1 mag (CR18),
equivalent to  N$_{\rm H}$ $\approx$ 2.2 $\times$ 10$^{22}$ cm$^{-2}$.
If the {\em Gaia} DR2 distance of 669$^{+50}_{-43}$  pc for V733 Cep is adopted then 
a more stringent upper limit log L$_{x}$(0.3-8 keV) $\leq$ 28.80 ergs s$^{-1}$
is obtained. The derived upper limits are sensitive to the assumed 
values of N$_{\rm H}$ and kT.  If the actual N$_{\rm H}$ is higher or
kT lower than assumed above, then the upper limits increase.
Despite its spectroscopic similarity to FU Ori, it is
much fainter in X-rays and so far undetected down to rather
stringent upper limits.

As shown in Figure 3, the X-ray luminosities of RNO 1B and 1C are comparable 
to accreting classical T Tauri stars (Preibisch et al. 2005; Telleschi et al. 2007)
despite the fact that FUors have much higher bolometric luminosities.
At our assumed distance of 930 pc, the results of Gramajo et al. (2014)
give L$_{bol}$ = 527  L$_{\odot}$ (RNO 1B) and 646  L$_{\odot}$ (RNO 1C).
These  L$_{bol}$ values are quite high, even for FUors (CR18). 
It is thus clear that high L$_{bol}$ for FUors does not necessarily translate
into high L$_{x}$. By comparison, a general trend for an increase in
L$_{x}$ with L$_{bol}$ is seen in classical T Tauri stars, albeit with 
large scatter in L$_{x}$ (Fig. 3). Since  L$_{bol}$ in FUors is dominated
by a luminous accretion disk, any possible dependence of L$_{x}$ on
the luminosity of the central star (L$_{*}$) will be difficult to 
establish without reliable estimates of  L$_{*}$.

Similarly, a correlation with stellar mass L$_{x}$ $\propto$ M$_{*}$ exists 
for classical T Tauri stars (Preibisch et al. 2005; Telleschi et al. 2007),
but no such correlation has yet been established for FUors since their
stellar masses are not well known.  There is even disagreement as to
whether the prototype  FU Ori is a subsolar mass star (Zhu et al. 2007)
or a more massive object (Herbig et al. 2003). But in a few cases, 
pre-outburst optical and near-IR observations show that the progenitor
was a young late-type star. For example, HBC 722 (= LkH$\alpha$ 188-G4 
= V2493 Cyg) was classified as a K7-M0 star by Cohen \& Kuhi (1979), 
implying a subsolar mass progenitor.

\subsection{Comments on FUor X-ray Emission}

Since some FUors are detected as luminous X-ray sources whereas
others are undetected down to rather stringent upper limits (e.g. V733 Cep),
we would like to know what factor(s) ultimately govern their X-ray properties.
Since the stars themselves are generally heavily-obscured, we lack sufficient 
data about stellar properties (e.g. mass, stellar luminosity, rotation period, 
magnetic field strength, accretion rate) to answer this question. However,
their bolometric luminosities have been determined and as noted above
the L$_{x}$ $\propto$ L$_{bol}$ relation seen in classical T Tauri stars 
does not hold for FUors. 

Previous observations  show that the {\em detected} X-ray emission 
of FUors cannot be due entirely to accretion shocks. Even though their
accretion rates are as high as 
$\dot{M}_{acc}$ $\gtsimeq$ 10$^{-5}$ M$_{\odot}$ yr$^{-1}$ during 
outbursts, accretion shock emission is expected to produce only
cooler X-ray plasma at characteristic temperatures T$_{shock}$ $\sim$ 1 - 2 MK
(kT$_{shock}$ $\sim$ 0.1 - 0.2 keV) for plausible infall speeds of a 
few hundred km s$^{-1}$ (Skinner et al. 2009). 
A similar conclusion  holds for shocked winds 
or jets assuming terminal speeds of a few hundred km s$^{-1}$ (Raga et al. 2002).
However, it is worth noting that soft accretion shock emission from very cool
plasma would be difficult to detect when viewed through the high absorption of some FUors.
Such objects include RNO 1C whose near-IR colors (Fig. 4) and high
A$_{\rm v}$ estimates (Sec. 1) indicate heavy reddening.
Furthermore, additional X-ray absorption 
from dust-depleted gas that is not accounted for by A$_{\rm v}$
may be present. This is probably the case for FU Ori which
shows X-ray evidence for cold gas near the star in the form
of fluorescent Fe I X-ray emission (Skinner et al. 2006). 
Another example is HBC 722 which showed apparently variable
circumstellar X-ray absorption by dust-depleted gas in
early outburst (Liebhart et al. 2016).

Even though the hot X-ray plasma detected in FUors cannot be 
attributed to accretion shocks, FUor X-ray properties might
be affected by accretion. This could occur if enhanced 
accretion during optical/IR outbursts alters the magnetic 
field topology or stellar structure, as has been suggested 
for accreting T Tauri stars (Preibisch et al. 2005) and the
eruptive young star V1118 Ori (Audard et al. 2010).

Fits of the  X-ray component detected in FUors with thermal 
plasma models typically 
give  kT $\gtsimeq$ 3 keV (T $\gtsimeq$ 35 MK). Our fits of 
the low-count RNO 1C {\em Chandra} spectrum are consistent 
with such high temperatures. In the case of FU Ori, the
X-ray count rate in the hard 2-8 keV band varied on a 
timescale of less than one day (Skinner et al. 2010). 
Such variable hard emission clearly points
to magnetically-controlled processes. The cause of the variability is 
not known, but the $<$1-day  timescale is suggestive of magnetic-reconnection 
flares. Even so, the factor of $\sim$2 variability in the hard-band count
rate of FU Ori is rather modest compared to the powerful impulsive X-ray
flares that have been detected in some T Tauri stars and class I protostars.

X-ray monitoring of FUors in the time domain is still quite limited.
Long-term X-ray monitoring is needed to search for powerful impulsive
X-ray flares and for evidence of periodic or quasi-periodic emission.
Periodic emission could be induced by stellar rotation and would provide
insight into poorly-known rotation periods, as has been
demonstrated for the protostar V1647 Ori (Hamaguchi et al. 2012).
Orbital X-ray modulation could also be present if some FUors are
close binaries, as suggested by Reipurth \& Aspin (2004).

\section{Summary}

{\em Chandra} observations of the close pair of similar FUor-like stars RNO 1B and 1C
reveal faint X-ray emission with intrinsic luminosities log L$_{x}$ = 29.64 -
29.83 ergs s$^{-1}$, making them the faintest X-ray detections among FUors to date.
Their low L$_{x}$ in combination with high L$_{bol}$ shows
that the L$_{x}$ $\propto$ L$_{bol}$ relation present in classical
T Tauri stars does not carry over to FUors.
Their X-ray spectral properties are not well-constrained due to the low number
of detected counts but our analysis suggests high plasma temperatures
kT $\gtsimeq$ 3 keV and high absorption N$_{\rm H}$ $\gtsimeq$ 10$^{22}$ cm$^{-2}$, 
the latter being consistent with existing A$_{\rm v}$ estimates. 
The high plasma temperature implies that the detected X-ray
emission is due to magnetic processes, not accretion shocks or shocked winds.

\smallskip
\smallskip

Support for this work was provided by {\em Chandra} X-ray Center award
GO 8-19004X.

\vspace{5mm}
\facilities{Chandra(ACIS)}

\vspace{5mm}
\software{CIAO (Fruscione et al. 2006),
          XSPEC (Arnaud 1996)}

\clearpage
                                                                                                                                                               
\end{document}